\documentstyle[11pt,aaspp]{article}
\newcommand{\BE}{\begin{equation}}
\newcommand{\BEAL}{\begin{eqnarray}}
\newcommand{\EE}{\end{equation}}
\newcommand{\EEAL}{\end{eqnarray}}
\def\la{L_a}
\def\lam{L_a^{-1}}
\def\lb{L_b}
\def\lbm{L_b^{-1}}
\def\wt{\tilde W}
\def\wtm{\tilde W^{-1}}
\def\wm{W^{-1}}

\def\l{\lambda}

\def\Ls{L^*}
\def\ls{\l^*}
\def\pk{P^{(k)}}

\begin{document}
\title{Some more exact results concerning multifield moduli of
 two-phase composites}
\author{ Mordehai Milgrom}
\affil{Department of Condensed-Matter
 Physics, Weizmann Institute of Science  76100 Rehovot, Israel}
\begin{abstract}
 \cite{c96} has recently shown how the response matrix of a  two-phase
composite can be written as certain linear combinations of products of
the component matrices. We elaborate on Chen's expansions by deriving
them in a different way, which a. shows them in a different light, and
b. permits us to generalize them further. As an application of our results
 we find exact microstructure-independent
 relations between the moduli of the two components and
those of any composite. The body of these relations is
 equivalent to the compatibility relations of \cite{msg89}, but
they are cast in a rather different form, which has certain advantages.
As an example, we show how any
 modulus of an arbitrary two-phase composite can be written in closed
 form as a linear combination of any other $n$
 of its moduli, with coefficients that depend
only on the component moduli, but not on the volume fractions, or
the microstructure.
\end{abstract}
\section{Introduction}
\par
 Our discussion concerns the response of a medium to many coupled fields, such
 as in the thermoelectric phenomenon, in the magneto-electric effect, or in
 coupled, multispecies diffusion. We concentrate of the response properties
 of a composite--whose multifield linear-response matrix
is $\Ls$--made of two components with response matrices $\la,~\lb$.
\cite{c96}, using a matrix-decomposition scheme,
 has recently shown that one can write
e.g.
\BE \Ls=a_0 \lb+a_1\la+a_2\la\lbm\la+...+
a_{n-1}\la\lbm\la...\lbm\la, \label{i} \EE
with $\la$ appearing $n-1$ times in the last term, were $n$ is the
number of coupled fields involved. Chen derives several
 other similar expansions, and there are similar expansions for
$(\Ls)^{-1}$. The expansion coefficients
are functions of the detailed makings of the composite--viz. the volume
fractions, and the microstructure.
\par
 We rederive and generalize Chen's expansions in \S 2, and determine their
coefficients in \S 3. In \S 4 we present a new, determinant  form
for the compatibility relations of \cite{msg89}.

\section{Rederivation and generalization of Chen's expansions}
\par

\par
Following \cite{msg89} we take $W$ to be a (non-singular) matrix that
diagonalize $\la$ and $\lb$ simultaniously by a congruence
transformation, such that
\BE W\la\wt=\l,~~~W\lb\wt=I,  \label{diag} \EE
where $\l=diag(\l_1,...,\l_n)$ is diagonal, and $I$ is the unit matrix.
Such a matrix always exists as $\la,~\lb$ are symmetric and positive
 definite. As explained in \cite{msg89}, the physics of the problem
dictates that the response matrix, $\Ls$, of any
composite material of the two components is also digonalized by $W$:
\BE W\Ls\wt=\ls=diag(\l^*_1,...,\l^*_n).  \label{lstar} \EE
\par
Consider now the series of $n\times n$ matrices $\pk$, for $k$ an
arbitrary integer, such that $\pk$ is transformed by $W$ into $\l^k$--the
$k$th power of $\l$:
\BE  \pk\equiv\wm\l^k\wtm. \label{dip} \EE

It is easy to see from eq.(\ref{diag})
 that $P^{(0)}=\lb,~P^{(1)}=\la$, and, in general,	 
\BE \pk=\lb(\lbm\la)^k.   \label{pow} \EE	
Thus, for positive $k$ $\pk=\la\lbm\la...\lbm\la$, 
 and for negative $k=-\vert k\vert$ $\pk=\lb\lam\lb...\lam\lb$, 
 with $\la$, or $\lam$ appearing $\vert k\vert$ times.
 If one interchanges the role of $\la$
 and $\lb$ ($W$ is than changed): $\pk(\la,\lb)=P^{(-k+1)}(\lb,\la)$.
\par
Let $N$ be the number of different $\l_i$s, than it is clear that:
\par
a. Any set of more than $N$ integer
 powers of the matrix $\l$ are linearly dependent
(because each can be viewed as a vector of length $N$).
\par
b. Any set of $N$ different (integer) powers, $\l^{k_1},...,\l^{k_N}$
are linearly independent. This is because the determinant of the matrix
\BE M(\vec\l,\vec k)\equiv\left(\matrix {\l_1^{k_1}&.&.&\l_1^{k_N}\cr
.&.& &.\cr
.& &.&.\cr
\l_N^{k_1}&.&.&\l_N^{k_N}\cr }\right) \label{dd}\EE
does not vanish unless at least two of the $\l_i$ are equal (for the
case of consecutive powers this is just proportional to the Vandermonde
determinant--see below)(see e.g. \cite{a49}).
\par
c. Any $n\times n$ diagonal matrix, with the same
 equalities among its element
 as in $\l$, can be written as a linear combination of any
set of $N$ different powers of $\l$.
Property c. is satisfied, in particular, by $\ls$ for any composite,
by its inverse, or, for that matter, by any integer power of it.
\par
For the sake of clarity we assume from now on that $\l$ is non-degenerate
i.e. $N=n$; all we say below goes though straightforwardly for $N< n$.
\par
Properties a.-c. are carried through by transforming back with
  $W$, whereby $\l^k$ goes back to $\pk$,
$f(\l)$ is carried back to $f(\la\lbm)\lb=\lb f(\lbm\la)$, $\ls$ goes to $\Ls$,
$(\ls)^{-1}$ goes to $\lb(\Ls)^{-1}\lb$, $(\ls)^m$ ($m$ positive) goes to
 $\Ls\lbm\Ls...\lbm\Ls$ ($\Ls$ appearing $m$ times), etc.
We thus have
\par
a'. Any set of more than $n$ of the $\pk$ matrices are linearly
 dependent.
\par
 b'. Any set of $n$ different $\pk$s are linearly independent.
\par
c'. Any of the matrices, such as $\Ls$, $\lb(\Ls)^{-1}\lb$, etc.
can be written as a linear combination of any $n$ of the $\pk$s
(any $N$ when $N<n$).
\par
So, for example, for every choice of different powers there are coefficients
$a_i$ such that
\BE \Ls=\sum_{i=1}^n a_i P^{(k_i)}. \label{gat} \EE
Chen's expansions are special cases of eq.(\ref{gat}) [and an analogous
 one for $\lb(\Ls)^{-1}\lb$] with consecutive values of $k_i$:
$k=k_0,k_0+1,...,k_0+n-1$, and special values of $k_0$. For example,
eq.(\ref{i}) is obtained when expanding $\Ls$ with $k_0=0$. Expanding
 $\lb(\Ls)^{-1}\lb$ using $k_0=-n+1$ gives
\BE (\Ls)^{-1}=b_0\lbm+b_1\lam+b_2\lam\lb\lam+...+b_{n-1}\lam\lb...\lb\lam,	
\label{man} \EE
another of Chen's expansions.

\par
Obviously, instead of using the $\pk$s as expansion basis, we could use
the response matrices of different composites of the same two composites,
or different space-space moduli matrices of anisotropic composites of the
same components. This will give relations between different composites
[for more details see the discussion in \cite{msg89}].

\section{The expansion coefficients}
\par

\par
The expansion coefficients $a_i$ depend on the exact nature of the
composite. It is sometimes useful to express these coefficients
in terms of $\l_i$ and $\ls_i$, which \cite{c96} does using a matrix
decomposition theorem. In the present formalism the $a_i$ are just
the coefficients of the linear expansion of the $n$-vector $\vec\ls$,
  made of
the diagonal of $\ls$, in the $n$ $n$-vectors made of the diagonals
of $\l^{k_i}$.
The vector $\vec a=(a_1,...,a_n)$ is thus the solution of the
linear equations
\BE M(\vec\l,\vec k)\vec a=\vec \ls,  \label{lol} \EE
where $M$ is the matrix defined in eq.(\ref{dd}).
[Their ilk are treated under `alternant matrices' in \cite{a49} (p. 111)]
\par
The case of consecutive $k$s lands itself to further simplification:
When $\vec k=(k_0,k_0+1,...,k_0+n-1)$ eq.(\ref{lol}) can be written as
\BE V(\vec\l)\vec a^{(k_0)}=\vec\delta^*,  \label{kpo} \EE
where $V$ is the Vandermonde matrix:
\BE V(\vec\l)\equiv\left(\matrix {1&\l_1&\l_1^2&.&.&\l_1^{n-1}\cr
1&\l_2&\l_2^2&.&.&\l_2^{n-1}\cr
.&.&.&.& &.\cr
.&.&.& &.&.\cr
1&\l_n&\l_n^2&.&.&\l_n^{n-1}\cr }\right), \label{qq}\EE
and $\delta^*_i\equiv \ls_i/\l_i^{k_0}$.
Thus
\BE a^{(k_0)}_i=\sum_j (V^{-1})_{ij}\ls_j/(\l_j)^{k_0}. \label{sok} \EE
The elements of $V^{-1}$ can be written in closed form:
\BE (V^{-1})_{ij}=(-1)^{i+j}\vert V_n\vert^{-1}\vert V_{n-1}^j\vert
S_{n-i}(\l_1,...,\l_{j-1},\l_{j+1},...,\l_n). \label{mata} \EE
Here, $\vert V_n\vert$ is the Vandermonde determinant for the $n$ $\l_i$,
 $\vert V_{n-1}^j\vert$ is the same for the $n-1$ $\l$s excluding $\l_j$,
 and $S_{n-i}$ is the elementary symmetric polynomial of degree $n-i$
 in its variables (here $\l_1,...,\l_{j-1},\l_{j+1},...,\l_n$); it is
 simply the sum of all possible products of $n-i$ subsets of different
 variables. So, for example, $S_0=1,~S_1=\sum_{k\not =j} \l_k$,
 and $S_{n-1}=\l_1\l_2...\l_{j-1}\l_{j+1}...\l_n$.			
 All this gives
 the expressions \cite{c96} obtains for the special values of
$k_0$ corresponding to his expansions.

\section{A new form of the compatibility relations}
\par
The fact underlying many an exact relation for
two-phase composites is that
 only $n$ numbers are needed to specify all the
the $n(n+1)/2$ moduli of a response matrix. Thus, the off-diagonal elements of 
 eq.(\ref{lstar}) can be viewed as $n(n-1)/2$ linear relations in the elements
 of $\Ls$ with coefficients involving only the component moduli (through $W$).		
 A given set of the
 compatibility relations of \cite{msg89}
 is another, and more useful, set of
 $n(n-1)/2$ relations between the moduli. In Chen's expansions and all
 their generalization discussed above [eq.(\ref{gat})] the
 $n$ structure-dependent
 parameters are the expansion coefficients. While all these sets of
 constraints are equivalent, each has its advantages. We now proceed to
derive yet other manifestations of the same constraints, which have their
own advantages.
\par
Let $A^1,...,A^{n+1}$ be any $n+1$ matrices that become diagonal when
transformed with $W$.
 They can be any of the $\pk$s, the response matrices of any
isotropic composite of the two components, the space-space response
matrix of an anisotropic composite,
 or appropriate products of such matrices
of the form $M_1~M_2^{-1}~M_3...M_{m-1}^{-1}~M_m$, where the $M$s are
diagonalized by $W$.
 The arguments of
\S 1\ show that these $n+1$ matrices are linearly dependent. Thus,
 if we choose any set of $n+1$ matrix slots: $r_1s_1,...,r_{n+1}s_{n+1}$,
the $n+1$ vectors
 $ \vec a_i=[A^i_{r_1s_1},...,A^i_{r_{n+1}s_{n+1}}], 
  ~~(1\le i\le n+1)$, are linearly dependent:
\BE \left\vert\matrix{A^1_{r_1s_1}&.&A^1_{r_{n+1}s_{n+1}}\cr
.&.&.\cr A^{n+1}_{r_1s_1}&.&A^{n+1}_{r_{n+1}s_{n+1}}}
\right\vert=0. \label{am} \EE
Take, for example, for the above $n+1$ matrices
 any $n$ different $\pk$s with $k=k_1,...,k_n$, and the response
matrix, $\Ls$, of some isotropic composite.
Then eq.(\ref{am}) tells us that
 \BE \left\vert\matrix {\Ls_{r_1s_1}&\Ls_{r_2s_2}&.&.&\Ls_{r_{n+1}
s_{n+1}}\cr
 P^{(k_1)}_{r_1s_1}&P^{(k_1)}_{r_2s_2}&.&.&P^{(k_1)}_{r_{n+1}
s_{n+1}}\cr
.&.&.& &.\cr
.&.& &.&.\cr
 P^{(k_n)}_{r_1s_1}&P^{(k_n)}_{r_2s_2}&.&.&P^{(k_n)}_{r_{n+1}s_{n+1}}\cr
}\right\vert=0, \label {plat} \EE
for any choice of different matrix slots $r_i s_i$.
\par
 Equation (\ref{plat}) constitute a linear relation, with structure-independent coefficients, between any chosen set of $n+1$ composite moduli.
Any set of independent $n(n-1)/2$ such determinant relations is
equivalent to one matrix compatibility relation of \cite{msg89} of the
form
\BE \Ls\lam\lb=\lb\lam\Ls,  \label{haut} \EE
which embodies the same number of independent relations between the
elements of $\la,~\lb,$ and $\Ls$.
The advantage of the latter is that it involves only elements
of the products $\lam\lb$, and $\lb\lam$, while the former involves
higher-order products $\pk$. The advantage of the former
is that each of its relations involve only $n+1$ of the elements
of $\Ls$, and they can be chosen at will, while in the latter each
 relation involves generically $2n-1$ elements of $\Ls$, which cannot
be chosen at will.
 The new relations eq.(\ref{plat}) thus permit us to
 write any modulus
of $\Ls$ as a microstructure-independent linear combination of any
other $n$ of its moduli. (Clearly, if we replace $\Ls$ in eq.(\ref{plat})
 by another of the $\pk$s we get an identity, as the derivation of such a
relation involves no physical arguments.)
\par
 One interesting example is the choice
with $n$ of the slots being the diagonal ones, and the
$(n+1)$th some off-diagonal element $rs$. This gives a determinant relation
of the form

\BE \left\vert\matrix {\Ls_{rs}&l^*_1&.&.&l^*_n\cr
 P^{(k_1)}_{rs}&p^{(k_1)}_1&.&.&p^{(k_1)}_n \cr
.&.&.& &.\cr .&.& &.&.\cr
 P^{(k_n)}_{rs}&p^{(k_n)}_1&.&.&p^{(k_n)}_n \cr}\right\vert=0,
 \label{og} \EE
where $l^*_i$ are the diagonal coefficients of $\Ls$, and
 $p^{(k_i)}_j$ is the $j$th diagonal element of
$P^{(k_i)}$.
 This form permits us to express all the off-diagonal
moduli of any composite in terms of the diagonal moduli (and those
of the components):
\BE \Ls_{rs}= \sum_i \alpha_i^{rs}~l^*_i,~{\rm where}~\alpha_i^{rs}
\equiv{U_i^{(rs)}\over \vert P\vert}. \label{ex}\EE
Here, $\vert P\vert$ is the determinant of the diagonal elements
 $p^{(k_i)}_j$, and $U_i^{(rs)}$ are the appropriate cofactors,
and all are calculated from the component moduli. An off-diagonal
modulus of an arbitrary composite is then given by a universal linear
function of the diagonal moduli of this composite.
\par
 For the special case when only two coupled fields are involved \cite{msg89}		
were, in fact, able to write the single compatibility relation in determinant
form (in this case $2n-1=n+1$). Our present result is a generalization to the	
$n>2$ case.
\par 
Equation(\ref{ex}) can be useful, for instance, in the following way:
If all cross moduli are small and can be neglected to lowest order, one can 
find simultaneous bounds on the diagonal elements $l^*_i\approx\ls_i$: A region
 $B$ in the space of $(l^*_1,...,l^*_n)$.		
 Equation(\ref{ex}), with the left-hand side a parameter,
defines a family of hyperplanes in this space. Bounds on individual
 $\Ls_{rs}$ can then be found by identifying the two extreme hyperplanes of
 the family still touching the region $B$. Note that these would constitute		
 (approximate) bounds on {\it individual} cross moduli. In contrast, the bounds
 discussed by \cite{msl89}, and by \cite{c96}, are exact but involve
 combinations of many moduli.
\par
All our results are applicable also to polycrystals made of a uniaxial
single crystal as is evident from the discussion of \cite{msp89}.

\acknowledgements
I thank Tungyang Chen for permission to use and quote his results before
publication.

\end{document}